\documentstyle[twocolumn,epsfig,prb,aps]{revtex}

\def\cn{{ \rm cn}}
\def\sn{{ \rm sn}}
\def\dn{{\rm dn}}

\begin{document}
\wideabs{

\title{Universal relation between the dispersion curve and the ground-state correlation length in 1D antiferromagnetic quantum spin systems}
\author{K. Okunishi, Y. Akutsu$^1$, N. Akutsu$^2$ and T. Yamamoto$^3$}
\address{Department of Physics, Niigata University, Igarashi 2, Niigata 950-2181, JAPAN \\
$^1$Department of Physics, Osaka University, Toyonaka, Osaka 560-0043, JAPAN \\
$^2$Department of Engineering, Osaka Electro-Comunication University, Neyagawa Osaka 572-8530, JAPAN \\
$^3$Department of Physics,  Gunnma Unisversity, Kiryuu, Gunnma 376-0052, JAPAN 
}
\date{\today}

\maketitle 

\begin{abstract}
We discuss an universal relation  $\varepsilon(i\kappa)=0$ with ${\rm Re} \kappa=1/\xi$ in 1D quantum spin systems with an excitation gap, where  $\varepsilon(k)$ is the dispersion curve of the low-energy excitation and $\xi$ is the correlation length of the ground-state.
We first discuss this relation for integrable models  such as the Ising model in a transverse filed and the XYZ model. 
We secondly  make a derivation of the relation for general cases, in connection with the equilibrium crystal shape in the corresponding 2D classical system.
We finally verify the relation for the $S=1$ bilinear-biquadratic spin chain and $S=1/2$ zigzag spin ladder  numerically.
\end{abstract}


}


\section{introduction}

Recently, considerable attention  has been paid for the low-energy excitations of one-dimensional(1D) antiferromagnetic(AF) quantum spin chains  with an excitation gap.
In discussing the low-energy excitations of the spin gap systems, one of the most fundamental quantities is the dispersion curve $\varepsilon(k)$ of the elementary excitation.
We can then expect that $\varepsilon(k)$  reflects the nature of the ground state correlation function $G(n)\equiv\langle S^z_0S^z_n \rangle$, since the elementary excitation can be interpreted as a ``local defect particle'' in the ground-state.
Actually for the $S=1$ AF Heisenberg chain, a relation $v=\xi\Delta$ is known numerically, where $v$ is the spin wave velocity, $\xi$ is the correlation length and $\Delta$ is the excitation gap.\cite{SoAf}
However,  a full justification of such a relation has not been made yet and to find its generalized form is an interesting problem.

In this paper,  on the basis of the correspondence between the 1D-quantum and 2D-classical systems, we present an universal relation between the low-energy dispersion curve and the ground-state correlation length:
\begin{equation}
\varepsilon(i\kappa)=0 \quad {\rm with} \quad {\rm Re} \kappa =1/\xi, \label{disp}
\end{equation}
which holds for a wide class of 1D quantum spin chains.
A remarkable point in eq. (\ref{disp}), which has not been stressed so far,  is that the full dispersion curve should be taken into account properly.
In other words, the conventional relation $v=\xi\Delta$ {\it does not hold} for general cases; $v=\xi\Delta$ is valid only for the relativistic-free-fermion case, where the dispersion curve is $\varepsilon(k)=\sqrt{\Delta^2+v^2\bar{k}^2}$ with $\bar{k}=\pi-k$.
Although the form of eq.(\ref{disp}) looks quite natural on the assumption of the free quasi-particle theory,  the validity of such a free particle picture is a highly nontrivial problem in the strongly correlated systems.

In the following, we address the relation (\ref{disp}) both in analytical and numerical ways.
We firstly show that the relation is actually satisfied for integrable models, where we can handle the 1D-quantum and 2D-classical correspondence exactly.
This analysis of the integrable models provides an essential insight for general cases.
We secondly  make a derivation of the relation for the general cases, dealing with the Suzuki-Trotter decomposed 2D classical systems\cite{ST}.
By using density-matrix renormalization group(DMRG)\cite{White} and the exact diagonalization tequniches, we finally  demonstrate that the relation is accepted for the  $S=1$ bilinear-biquadratic(BLBQ) chain and the $S=1/2$ zigzag spin ladder, which exhibit characteristic dispersion curves associated with the  cusp singularity in the magnetization process.\cite{OHA,Gol,ladder,ftemp}

\section{exactly solvable cases}

Although a lot of important relations are established in the integrable systems\cite{Bax}, eq. (\ref{disp}) has not been discussed yet. Thus we start with analyzing some integrable lattice models.
As the simplest  but important example,  let us examine the 1D Ising model in a transverse field(ITF): $H=J\sum s^z_is^z_{i+1} - \Gamma \sum s^x_i$, where $\vec{s}_i$ is the spin-$1/2$ operator at $i$ th site and $\Gamma$ is the transverse field.
This model is solved with the Jordan-Wigner transformation, and then the dispersion curve is obtained as $\varepsilon_{\rm ising}(k)=\frac{1}{2}\sqrt{(J\cos k + 2\Gamma)^2 + (J\sin k)^2  }$.\cite{Pfeuty}
The correlation function can be also calculated straight forwardly to be $\xi^{-1}=\left| \ln (2J/\Gamma) \right|$.
 We can then confirm $\varepsilon_{\rm ising}(i/\xi)=0$ clearly.

More generally, we consider the XYZ model: $H=\sum_i J_x \sigma^x_i\sigma^x_{i+1} + J_y \sigma ^y_i\sigma^y_{i+1} + J_z \sigma^z_i\sigma^z_{i+1}$, where $\vec{\sigma}_i$ is the Pauli matrix at $i$-th site and the coupling constants are represented by the Baxter's elliptic-function parametalization with a modulus $l$: $J_x=\cn 2\zeta /\sn 2\zeta  $, $J_y=\dn 2\zeta/\sn 2\zeta$ and $J_z=1/ \sn 2\zeta$.\cite{Bax}
We consider only the AF region of the couplings. 
The dispersion curve of the spinon excitation is calculated exactly by Johnson, Krinsky and McCoy:\cite{JKM}
\begin{equation}
\varepsilon_{\rm XYZ}(k)=2J_z \frac{K_1\sn(2\zeta)}{K'}\sqrt{1-k_1^2 \cos^2 k},\label{xyzdisp}
\end{equation}
where $K'$ is the complete elliptic integral of the modulus $l$, and  $k_1$ is the modulus defined by  $K'_1/K_1\equiv\pi\zeta/K'$.
The correlation function of the XYZ chain is also calculated: 
\begin{equation}
G(n)=-\int\!\!\int d\phi_1d\phi_2 \rho(\phi_1,\phi_2)\left[k_2 \sn (\frac{K_2\phi_1}{\pi})\sn(\frac{K_2\phi_2}{\pi})\right]^n \label{8vcf}
\end{equation}
where $\rho(\phi_1,\phi_2)$ is a non-singular function and $k_2$ is another modulus defined by the elliptic integrals $K'_2/K_2\equiv 2\pi\zeta/K'$.\cite{JKM}
In order to compute the asymptotic form of the correlation function,  we apply the  saddle point method to eq. (\ref{8vcf}) and obtain $G(n)\sim k_2^n$.
Since the $\sigma^z$ operator creates two spinon excitations when it acts on the ground state wavefunction, the spinon-spinon correlation length is given by $\xi^{-1}=-\frac{1}{2}\ln k_2$. 
Then, we can see $\varepsilon_{\rm XYZ}(\frac{-i}{2}\ln k_2)=0$ easily, by using the Landen transformation.
From these, we have verified eq. (\ref{disp})  for the integrable models.

\section{general cases}

The realization  of eq. (\ref{disp}) for the integrable models can be attributed to the nature of the underlying 2D classical models, such as 8-vertex model.
An useful point in considering the 2D classical systems is that  the asymptotic behavior of the system in the time and space directions can be treated systematically; Particularly, it should be remarked that the anisotropic correlation length of the system can be associated with the equilibrium-crystal shape(ECS) of the interface via the Wulff construction: \cite{Wulff}
\begin{equation}
\gamma(\theta)= \frac{k_BT}{\xi(\theta)},
\end{equation}
where $\gamma(\theta)$ is the crystal shape at the angle $\theta\equiv \arctan(y/x)$ and $\xi(\theta)$ is the angle dependent correlation length.
For the integrable models,  the universal behavior of the ECS is shown exactly,  and then the eq. (\ref{disp}) for the corresponding 1D quantum spin systems can be explained  as one of the  resulting behaviors of the universal  ECS.

In order to see the connection of the ECS with the quasi-particle excitation, we represent the correlation length $\xi(\theta)$  as the asymptotics of the lattice Green's function: 
\begin{equation}
 \frac{1}{\xi(\theta)} = -\lim_{r\to\infty} \frac{1}{r}\ln {\cal  G}(x,y)
\end{equation}
where  ${\cal G}(x,y)$ is the lattice greens function and $r=\sqrt{x^2+y^2}$.  Throught the Fourier space pepresentation, we can write this lattice Green's function as 
\begin{equation}
{\cal G}(x,y)=\int_{-\pi}^{\pi} dk_x \int_{-\pi}^{\pi} dk_y \frac{e^{ik_x x + ik_y y }}{D(k_x,k_y)},
\label{lgf0}
\end{equation}
in which the asymptotic behavior in the real space is determined by the condition $D(k_x,k_y)=0$. \cite{Holzer,AA}
The correlation length and the dispersion curve in the 1D quantum system can be also  extracted from $D(k_x,k_y)$ (see below in detail).
In fact  $D(k_x,k_y)$ of the XYZ model is  calculated in Ref.\cite{Fujimoto}, where we can reduce the universal curve of the ECS to eq. (\ref{disp}).

In the framework of the Wullf construction, what we want to emphasize here is that there is the one-to-one correspondence between the ECS and the asymptotics of the lattice Green's function.
Since the ECS is the well-defined thermodynamic object\cite{ecscondition} and the Wulff construction does not require the integrability of the system, we can assume the existence of the lattice Green's function with $D(k_x,k_y)$ for a wide class of non-integrable models.
This implies that the asymptotic form of the lattice Green's function (\ref{lgf0}) can be  defined {\it not} by the perturbative argument of the  local excitation, {\it but by the thermodynamic quantity of the ECS, where the interaction effect of the system is fully taken into account}.

To analyze the general 1D quantum systems, we consider the dynamical correlation function $G(n,\tau)\equiv \langle S^z(n,\tau)S^z(0,0)\rangle_\beta$, where  $S^z(n,\tau)$ is the spin operator of the Heisenberg representation at position $n$ and imaginary time $\tau$, and $\langle \cdots \rangle_\beta$ means the thermal average at an inverse temperature $\beta$.
Our goal is to evaluate the asymptotic behavior of $G(n,\tau)$ in the zero-temperature limit.
For this purpose we describe  $G(n,\tau)$ by the lattice Green's function (\ref{lgf0}) on the Suzuki-Trotter transformed 2D classical system of the size $N\times 2M$:
\begin{equation}
{\cal G}(n, 2m)=\int_{-\pi}^{\pi} dk_x \int_{-\pi}^{\pi} dk_y \frac{e^{ik_x n + ik_y 2m}}{D(k_x,k_y)},
\label{lgf1}
\end{equation}
where $m\equiv M\tau/\beta$,  $N$ is the lattice size in the spatial direction, and $M$ is the Trotter number.
In the Trotter limit $M\to \infty$, we recover the original Green's function  $G(n,\tau)\simeq {\cal G}(n, 2m)$.
The function ${D(k_x,k_y)}$ should be determined when a model is given. 
For example, that of the ITF is given by $D(k_x,k_y)=\cosh 2K_x \cosh 2K_y -\sinh 2K_x \cos k_x - \sinh 2K_y \cos k_y$ with $K_x=\beta J/(4M)$ and $K_y=\frac{1}{2}\ln[\coth(\frac{\beta\Gamma}{M})]$.\cite{AA}
However, the explicit form of $D(k_x,k_y)$ is not required in the following discussion.

In eq. (\ref{lgf1}), we first consider the Trotter limit $M\to \infty$, where the dominant contribution to the $k_y$ integral comes from $k_y\sim 0$, since $e^{ ik_y 2m }$ is highly oscillating.
Substituting $k_y=\frac{\beta}{2M} \tilde{k_y}$, we can expand $D(k_x,k_y)$ as 
\begin{equation}
D(k_x,k_y)=D(k_x,0) +D_{yy}(k_x,0)(\frac{\beta}{2M}\tilde{k}_y)^2 + \cdots,
\label{dexpantion}
\end{equation}
where the first-order term of the $k_y$ derivative vanishes due to the Helmiticity of the original 1D spin chain.
Though $D(k_x,k_y)$ contains the Trotter number dependence, we find that $D(k_x,0) +D_{yy}(k_x,0)(\frac{\beta}{2M}\tilde{k}_y)^2$ has a proper limit  as $M\to \infty$, without the overall normalization factor, and then the higher-order terms with respect to $\tilde{k_y}$ can be neglected.
Thus we obtain 
\begin{eqnarray}
{\cal G}(n, 2m)= \int_{-\pi}^{\pi} dk_x F(k_x) \int_{-\infty}^{\infty} d\tilde{k}_y \frac{e^{ik_x n + i\tilde{k}_y \tau}}{d(k_x)+\tilde{k}_y^2}
\label{lgf2}
\end{eqnarray}
where $F(k_x)$ is a non-singular function and we have defined 
\begin{equation}
d(k_x)\equiv \lim_{M\to \infty}\frac{2MD(k_x,0)}{\beta D_{yy}(k_x,0)}. \label{dfunc}
\end{equation}
For the case of the ITF, we have  $d(k_x)=\frac{1}{4}(J^2+4\Gamma^2+4J\Gamma \cos k_x )$, and $F(k_x)=2M\Gamma/\beta $ explicitly.

We next consider the low-energy limit $\beta\to \infty$.
We then evaluate the ``long-distance behavior'' of eq. (\ref{lgf2})  with a fixed direction  $\tau/n \equiv \tan\theta$ in the limit $n,\tau \to \infty$.
Applying the saddle point method,  we obtain $G(n,\tau)\sim e^{-\kappa_xn-\kappa_y \tau}$, where  $\kappa_x$ and $\kappa_y$ satisfies the saddle point conditions:
\begin{eqnarray}
d(i\kappa_x)-\kappa_y^2=0 , \qquad \frac{2\kappa_y}{ \frac{\partial d(i\kappa_x)}{\partial (i\kappa_x)} }=\tan \theta .\label{saddlepoint}
\end{eqnarray}
Particularly for the case of the equal-time correlation function, we set $\kappa_y=0$($\theta=0$).
Then $G(n,\tau)$ reduces into $G(n)\sim e^{-\kappa_x n}$ with $d(i\kappa_x)=0$, in which the real part of $\kappa_x$ represents the inverse correlation length of the ground-state.

Let us here discuss the meaning of the function $d(k_x)$.
To see it,  we further  perform the $\tilde{k}_y$ integral in eq. (\ref{lgf2}) and then  obtain another expression of the Green's function:
\begin{eqnarray}
G(n,\tau)={\rm const}\int_{-\pi}^{\pi} dk_x  \frac{e^{ik_x n - \sqrt{d(k_x)}\tau}}{\sqrt{d(k_x)}}.\label{integratedcf}
\end{eqnarray}
On the other hand, the asymptotic behavior of the Green's function can be reduced into  $G(n,\tau)\sim \int  dk \hat{G}(k)e^{ik n - \varepsilon(k)\tau}$ in the low-energy limit, by taking account of the one-particle part of the excitation, where $\hat{G}(k)$ is the Fourier transform of $G(n)$.
Comparing the above expression with eq. (\ref{integratedcf}), we can see $\sqrt{d(k)}=\varepsilon(k)$, which is checked for the ITF easily.
Hence, we finally arrive at the formula (\ref{disp}).

\section{numerical results}


In the above derivation,  we have formally written down the function $D(k_x,k_y)$.
In this section,  we verify the relation (\ref{disp}) for  non-integrable models in numerical ways, since the analytic  calculation of  $D(k_x,k_y)$ is very difficult.
As a typical example of the non-integrable models, we investigate the $S=1$  BLBQ chain, which plays an important role to understand the Haldane systems. 
The Hamiltonian of the BLBQ chain is given by 
\begin{equation}
{\cal H}=\sum_i \left[ \vec{S}_i\cdot \vec{S}_{i+1} + \beta(\vec{S}_i\cdot \vec{S}_{i+1})^2 \right] ,   \label{blbq}
\end{equation}
where $\vec{S}_i$ is the $S=1$ spin operator at $i$-th site, and $\beta$ is the coupling constant of the  biquadratic term(not inverse temperature).
The BLBQ chain with $\beta=0$, $1/3$ and $1$ corresponds to the $S=1$ Heisenberg spin chain, the valence-bond-solid(VBS) chain\cite{AKLT} and the $SU(3)$ integrable chain\cite{SU3} respectively.

In connection with the relation (\ref{disp}), a remarkable feature of the BLBQ chain is that its dispersion curve and the ground-state correlation function exhibit various characteristic properties; 
Some extensive numerical studies of the ground-state correlation function claim that the VBS point is a disorder point, which is the onset of the incommensurate oscillation of $G(n)$ accompanying with a cusp singularity in $\xi$.\cite{Schol}
Appearance of Lifsitz point $\beta=\beta_L\simeq0.43$ is also reported, where the peak wavenumber of the static structure factor starts to shift as $\beta$ increases. \cite{Schol}
Moreover, at $\beta=\beta_c\simeq 0.40$, the shape of the dispersion curve changes into the double well structure,\cite{Gol} which is associated with the cusp singularity in the magnetization process.\cite{OHA}
Although a phenomenological explanation of this scenario is recently made by assuming a continuous-free-field theory,\cite{Fath2} a full justification  based on the lattice Hamiltonian (\ref{blbq}) has not been achieved yet.
We now demonstrate the our relation (\ref{disp}) for the BLBQ chain, which provides the background for the phenomenological theory.

The ground-state correlation function of the BLBQ chain can be investigated with the DMRG,\cite{Schol} while there is few systematic study of the whole shape of the dispersion curve for the sufficiently long chain.
For the purpose of computing the dispersion curve, we employ the DMRG  combined with the continued fraction method, where the excitation spectrum is extracted as a peak of the imaginary part of the dynamical correlation function expressed as the continued fraction.\cite{ddmrg1,ddmrg2}
The DMRG calculation is performed  for the BLBQ  chain of the length $L=240$ with the open boundary condition,  targeting up to the five Lanczos vectors generated form the initial state $\hat{S}^z_k|0\rangle$, where $\hat{S}^z_k\equiv \sum S^z_n e^{ikn}$.
In addition to this, we use the filtering technique to cut off the scattering at the boundaries, according to Ref.\cite{ddmrg2}. 
In FIG.1 we show the dispersion curves for $\beta=0$, $1/3$ and $0.6$, which are calculated with the maximum number of the retained bases $m=81$.
Clearly we can see the shape change of the dispersion curve induced by the incommensurate effect originating from the biquadratic term.
The dotted parts in the curves for $\beta=1/3$ and $0.6$ mean that the dispersion curves are screened by the continuum spectrum.

\begin{figure}[h]
\epsfig{file=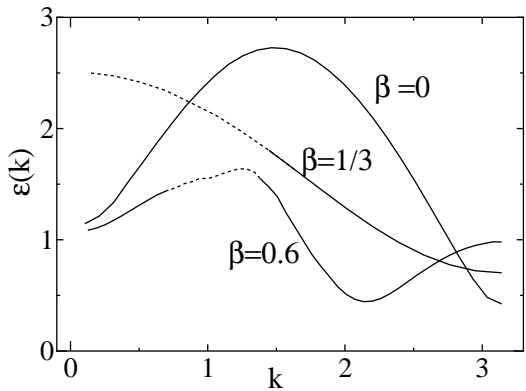,width=7cm}
\caption{Dispersion curves of the BLBQ chain for $\beta=0$, $1/3$, and $0.6$}
\end{figure}

In order to check eq. (\ref{disp}), we estimate the Fourier series expansion of the dispersion curve with a fitting.
Using the fitted curve, we determine the zero of the dispersion curve in the complex plane and compare the result with the correlation length obtained from a direct computation of $G(n)$.
Here let us recall the dispersion curve has the square-root form $\varepsilon(k)=\sqrt{d(k)}$.
Thus we can expect that the function $d(k)$ is fitted well rather than $\varepsilon(k)$:
\begin{eqnarray}
d(k)=\varepsilon(k)^2= \sum_{n=0}^N A_n\cos nk, \label{sqfit}
\end{eqnarray}
where $\{ A_n\}$ is fitting parameters and $N$ denotes the number of the parameters(cut off frequency).
For the Heisenberg point $\beta=0$,  the  fittings with $N=5$, $6$ and $ 7$ yield $\xi=0.590(5)$, $6.02(8)$ and $0.60(1)$,\cite{accuracy} which are converged with respect to $N$ and in good agreement with the known value $\xi=6.03$.
We also test the fitting: $\varepsilon(k) =  B_0 + B_1 \cos(k)+B_2 \cos 2k +\cdots$, and obtain  $\xi= 2.5(1)$, $2.8(2)$ and $3.2(2)$ for $N=6$, $7$ and $8$ respectively, which does not converge with respect to $N$. 
This slow convergence for $\varepsilon(k)$ may be essential, since $\varepsilon(i\kappa)=0$ corresponds to the blanch-cut singularity of the square-root function.
Hence, we have verified the $\varepsilon(i\kappa)=0$ with $\kappa=1/\xi-i\pi$, including the square-root form of the dispersion curve.
Here, it should be noted that we obtain  $\xi=6.00$  for the case of assuming the relativistic free fermion: $\varepsilon(k) = \sqrt{\Delta^2+v^2  \bar{k}^2}$ with $\Delta=0.410$ and $v=2.46$.

At the VBS point, the  fittings with $N=3$, $4$ and $5$  yield $\xi=0.92(1)$, $0.90(1)$ and $0.91(1)$ respectively,  which are consistent with the exact value $\xi=1/\ln 3=0.910\cdots$.
On the other hand, the relation $v=\xi\Delta$ with $\Delta=0.70$ and $\xi=0.91$ gives the velocity $v=0.64$, which does not agree with the correct value $v=0.85$ estimated by the curvature at the bottom of the dispersion curve.
Thus the conventional relation is not maintained in general cases.
For the VBS model, we can further  discuss $\varepsilon(\pi+i/\xi)=0$ exactly, based on the domain-wall picture of the excitation, where the dispersion curve is calculated as the Fourier transform of the domain-wall correlation function.\cite{Arov1,Fath1}
Since the asymptotic behavior of the domain-wall correlation function is determined by the overlap integral between the four degenerating VBS states, we can see the dispersion curve  contains the factor $5+3 \cos k$, which yields the zero of $\varepsilon(k)$ at $k=\pi+i\ln 3$.
In addition, we should remark that the solution of $d(k)=0$ is  given by the twofold root of $k=\pi+i\ln 3$, because $d(k)=\varepsilon(k)^2$.
This implies that the VBS model locates at the disorder point, since the imaginary part of the solution of  $d(k)=0$ starts to move from $\pi$ with making a complex-conjugate pair, as $\beta$ is increased.

For $\beta=0.6$,  the numerical data in the range $\pi/2<k\le \pi$ is well converged, where the imaginary part of the dynamical correlation function has sharp peaks. 
On the other hand, the curve in $0.6\pi<k<\pi/2$ is disturbed by the continuum spectrum.
However, we find that the main contribution to the zero of the $\varepsilon(k)$ comes from the range $\pi/2<k\le \pi$.
Actually,  we have obtained the sufficiently stable results: $\xi=3.79(5)$ and $q=0.678(1)\pi$ and $\xi=3.80(2)$ and $q=0.680(2)\pi $, by the fitting of eq. (\ref{sqfit}) with $N=7$ and $8$ respectively, where $q$ denotes the incommensurate period of the correlation function.
These values agree with $\xi=3.87$ and $q=0.678\pi$,  which is computed directly from the real-space  correlation function.
Thus the formula (\ref{disp}) is verified for the BLBQ chain, including the period of the incommensurate oscillation.


The similar behaviors of the dispersion curve and the correlation function can be expected in a class of the frustrated spin ladder systems.\cite{ladder,ladder1,mobius}
In the remaining part of this section, we further test the relation  (\ref{disp}) for  the $S=1/2$ zigzag spin ladder in the dimer gap phase, which also shows the characteristic behaviors of the dispersion curve and the correlation function.\cite{ladder,mobius}
The Hamiltonian of the zigzag ladder is given by
\begin{equation}
{\cal H}= \sum_i \left[ \vec{S}_i\cdot \vec{S}_{i+1} + \alpha \vec{S}_i\cdot \vec{S}_{i+2} \right],
\end{equation}
where $\vec{S}$ is the $S=1/2$ spin operator.
The zigzag ladder is in the dimer gap phase for $\alpha>0.2411$, and $\alpha=0.5$ is called Majumdhar-Gosh point, where the ground state is exactly represented as the dimer singlet state.
For $\alpha>0.54$,  the spinon dispersion curve has the double well structure,  in accordance with the incomennsurate oscillation of the correlation function.
The magnetization curve has also a cusp singularity in this region of $\alpha$.\cite{ftemp,mobius}

An important feature of the zigzag ladder is that the elementary excitation is described by the topological excitation of $S=1/2$,  which is called spinon.
Recently, the single spinon excitation of the zigzag ladder has been extracted by using the exact diagonalization method with the M\"obius boundary condition,\cite{mobius} (See Fig.2)
although it has not been obtained for the system with the periodic boundary condition where the excitation is basically described by the combination of the two spinons.
From the spinon dispersion curve calculated with the M\"obius boundary condition,  we estimate the square of the dispersion curve $d(k)$ with the fitting of (\ref{sqfit})  and solve  $d(i\kappa)=0$ numerically. 
We then compare the results with the correlation length computed with the DMRG method.

\begin{figure}[h]
\epsfig{file=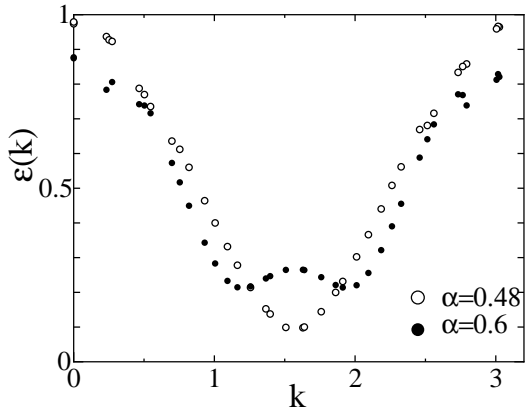,width=7cm}
\caption{Single spinon dispersion curves of the zigzag spin ladder for $\alpha=0.48$(open circle) and $0.6$(solid circle).}
\end{figure}

We here consider the zigzag ladder of $\alpha=0.48$ and $\alpha=0.6$ typically, whose properties are similar to the BLBQ chain of $\beta=0$ and $\beta=0.6$ respectively. 
For $\alpha=0.48$,  the period of the correlation function is commensurate and thus the bottom of the dispersion curve is located at $k=\pi/2$.
In fact, the result of $d(i\kappa)=0$ yields $\xi=5.71(6)$, which is consistent with  the correlation length obtained with the DMRG: $\xi=5.68(7)$.
For $\alpha=0.6$, the single spinon dispersion curve has a double well structure around $k=\pi/2$, and then the solution of  $d(i\kappa)=0$ yields $\xi=4.35(20)$ and $q= 0.368(5) \pi$.
On the other hand, the asymptotic behavior of the correlation function computed  with the DMRG is $\xi=4.26(20)$ and  $q=0.367(2)$, which is also in good agreement with the results of the relation (\ref{disp}).
We have therefore verified  the relation (\ref{disp}) for the $S=1/2$ zigzag spin ladder, including the period of the incommensurate oscillation of the correlation function.

\section{summary and discussion}

We have presented the relation (\ref{disp}), which connects the zero of the dispersion curve with the correlation length of the ground-state.
For the integrable systems, we have verified the relation by the exact expressions of the dispersion curve and the correlation length.
We have next derived the relation (\ref{disp}) for general cases, based on the Suzuki-Trotter transformed 2D classical models, where the connection to the universal crystal shape of the equilibrium interface plays an important role.
To check eq. (\ref{disp}) in the non-integrable systems,  we have computed the dispersion curve of the BLBQ chain, using the DMRG method combined with the continued fraction method.
We have also investigated eq. (\ref{disp}) for the $S=1/2$ zigzag spin ladder in the dimer gapped phase. 
Then we have shown that the relation (\ref{disp}) explains the asymptotic behavior of the correlation function including the incommensurate oscillation.

Since the argument in this paper is founded on the equilibrium crystal shape in the corresponding classical system,  the validity of the relation (\ref{disp}) is independent of such special properties of the system  as the integrability, the anisotropy of the system(full-$SU(2)$ symmetry), the spin magnitude of the excitation particle, and so on.
Indeed, the qasi-particle excitations considered in this paper includes various types of the excitations.
We can, therefore, conclude that the relation (\ref{disp}) is  hold for a wide class of quasi-1D quantum systems.
Recently,  some interesting problems between the correlation function and the dispersion curve are also expected in a class of frustrated spin systems or correlated electron systems.\cite{ladder1} 
When analyzing such systems, we believe that the relation (\ref{disp}) plays an important role.

\acknowledgements

The authors would like to thank T. Nishino and N. Maeshima for helpful comments.
K.O. was supported by the Japan Society for the Promotion of Science.



\begin{thebibliography}{99}
\bibitem{ecscondition} A counter example is the spin chain in the saturation field, where the ECS does not draw the closed curve. However, such a situation is rather extream.
\end{thebibliography}
\end{document}